\documentclass[prd,superscriptaddress,amsfonts,amssymb,amsmath,showpacs,onecolumn]{revtex4-2}
\usepackage{bm}
\usepackage{amsfonts}
\usepackage{latexsym}
\usepackage[latin1]{inputenc}
\usepackage{graphicx}
\usepackage{amsmath}
\usepackage{palatino}
\usepackage{ragged2e}
\usepackage{mathpazo}
\usepackage{textcomp}
\usepackage{float}
\usepackage{booktabs}
\usepackage{dcolumn}
\usepackage{multirow}
\usepackage{hyperref}
\hypersetup{colorlinks,citecolor=blue}
\usepackage{amsmath}
\usepackage{xcolor}
\usepackage{orcidlink}
\usepackage[caption=false]{subfig}
\usepackage{commath}
\def\jnl@style{\it}
\def\aaref@jnl#1{{\jnl@style#1}}

\def\aaref@jnl#1{{\jnl@style#1}}

\def\aj{\aaref@jnl{AJ}}                   % Astronomical Journal
\def\apj{\aaref@jnl{ApJ}}                 % Astrophysical Journal
\def\apjl{\aaref@jnl{ApJ}}                % Astrophysical Journal, Letters
\def\apjs{\aaref@jnl{ApJS}}               % Astrophysical Journal, Supplement
\def\apss{\aaref@jnl{Ap\&SS}}             % Astrophysics and Space Science
\def\aap{\aaref@jnl{A\&A}}                % Astronomy and Astrophysics
\def\aapr{\aaref@jnl{A\&A~Rev.}}          % Astronomy and Astrophysics Reviews
\def\aaps{\aaref@jnl{A\&AS}}              % Astronomy and Astrophysics, Supplement
\def\mnras{\aaref@jnl{Mon.~Not.~Roy.~Astron.~Soc.}}             % Monthly Notices of the RAS
\def\prd{\aaref@jnl{Phys.~Rev.~D}}        % Physical Review D
\def\prc{\aaref@jnl{Phys.~Rev.~C}}  % Physical Review C
\def\prl{\aaref@jnl{Phys.~Rev.~Lett.}}    % Physical Review Letters
\def\qjras{\aaref@jnl{QJRAS}}             % Quarterly Journal of the RAS
\def\skytel{\aaref@jnl{S\&T}}             % Sky and Telescope
\def\ssr{\aaref@jnl{Space~Sci.~Rev.}}     % Space Science Reviews
\def\zap{\aaref@jnl{ZAp}}                 % Zeitschrift fuer Astrophysik
\def\nat{\aaref@jnl{Nature}}              % Nature
\def\aplett{\aaref@jnl{Astrophys.~Lett.}} % Astrophysics Letters
\def\apspr{\aaref@jnl{Astrophys.~Space~Phys.~Res.}} % Astrophysics Space Physics Research
\def\physrep{\aaref@jnl{Phys.~Rep.}}      % Physics Reports
\def\physscr{\aaref@jnl{Phys.~Scr}}       % Physica Scripta
\def\commat{\aaref@jnl{Comm.~Math.~Phys.}}              % Communications in Mathematical Physics
\def\science{\aaref@jnl{Science}}               % Science
\def\cqg{\aaref@jnl{Classical Quant.~Grav.}}            % Classical and Quantum Gravity
\def\jpcs{\aaref@jnl{JPCS}}                                     % Journal of Physics Conference Series
\def\ijmpd{\aaref@jnl{Int.~J.~Mod.~Phys.~D}}                    % International Journal of Modern Physics D
\def\grg{\aaref@jnl{Gen.~Relat.~Gravit.}}               % General Relativity and Gravitation
\def\rpp{\aaref@jnl{Rep.~Prog.~Phys.}}          % Reports on Progress in Physics
\def\npa{\aaref@jnl{Nucl.~Phys.~A}}        % Nuclear Physics A
\def\lrr{\aaref@jnl{Living Rev.~Rel.}}                   % Living reviews in relativity
\def\jcap{\aaref@jnl{J.~Cosmology Astropart.~Phys.}}    % Journal of cosmology and astroparticle physics
\def\rmp{\aaref@jnl{Rev.~Mod.~Phys.}}   %Reviews of modern physics
\def\epjc{\aaref@jnl{Eur.~Phys.~J.~C}}

\allowdisplaybreaks[1]
\begin{document}

\color{black}       %% For one column

\title{Baryogenesis in $f(R,L_m)$ gravity}

\author{Lakhan V. Jaybhaye\orcidlink{0000-0003-1497-276X}}
\email{lakhanjaybhaye@gmail.com}
\affiliation{Department of Mathematics, Birla Institute of Technology and
Science-Pilani,\\ Hyderabad Campus, Hyderabad-500078, India.}

\author{Snehasish Bhattacharjee\orcidlink{0000-0002-7350-7043}}
\email{snehasish@astro.ncu.edu.tw}
\affiliation{Institute of Astronomy, National Central University, 32001 Taoyuan, Taiwan}
\affiliation{Department of Particle Physics and Astrophysics, Weizmann Institute of Science, 76100 Rehovot, Israel}

\author{P.K. Sahoo\orcidlink{0000-0003-2130-8832}}
\email{pksahoo@hyderabad.bits-pilani.ac.in}
\affiliation{Department of Mathematics, Birla Institute of Technology and
Science-Pilani,\\ Hyderabad Campus, Hyderabad-500078, India.}

%%%%%%%%%%%%%%%%%%%%%%%%%%%%%%%%%%%%  DATE  %%%%%%%%%%%%%%%%%%%%%%%%%%%%%%%%%%%%
\date{\today}

\begin{abstract}

This paper aims to recreate the gravitational baryogenesis epoch in the framework of the $f(R,L_m)$ theory of gravity, where $R$ and $L_m$ are  the curvature scalar and the matter Langragian, respectively. In particular, we examine the model, $f(R,L_m) = \frac{R}{2} + L_m ^{\alpha} + \zeta$, under the supposition that the universe is saturated with dark energy and perfect fluid, with a non-zero baryon to entropy ratio during a radiation dominance era. We confine the model with the gravitational baryogenesis scenario, emphasizing the appropriate values of model parameters compatible with the baryon-to-entropy ratio observation data.
Our study found that  $f(R,L_m)$ gravity can considerably and steadily make a contribution to the phenomenon of gravitational baryogenesis.
\end{abstract}

\maketitle

\section{ Introduction}\label{sec1}
\justify 

Even before cosmology became an independent research division, one of the unanswered questions was the abundance of matter over antimatter in our Universe. Significant observational evidence, like Big-Bang Nucleosynthesis (BBN) \cite{Burl} and the Cosmic Microwave Background (CMB) \cite{Benn}, has strongly indicated that matter predominates over antimatter in the universe. This superiority is known as baryogenesis.
Recent research suggests that the asymmetrical correlation between matter and antimatter originally came at the beginning of the cosmos. However, the true origin of baryon asymmetry (BA) remains a mystery that requires further investigation.\\
Numerous theories have emerged to solve this mystery of BA by evaluating interactions in the primitive universe that go beyond the standard model, some of which are Affleck-Dine baryogenesis \cite{Stew,Yama,Akit}, spontaneous baryogenesis \cite{Taka,Bran,Simo} electroweak baryogenesis \cite{Trod,Morr}, grand unified theories (GUTs)\cite{Kolb}, baryogenesis of thermal and black hole evaporation \cite{Dolg}, these multiple baryogenesis contexts discuss how this universe could have more matter than antimatter during the matter or radiation epoch. The gravitational baryogenesis process employs one of the Sakharov criteria \cite{Sakh} proposed that the baryon asymmetry can be produced by three necessary conditions: (1) processes that violate baryon number, (2) Charge (C) and ChargeParity (CP) violation, and (3) Out of thermal equilibrium interactions. The key component is a CP-violating interaction stipulated by coupling between the baryon matter current $J^\mu$ and the derivative of the Ricci scalar curvature $R$, in the form

\begin{equation}\label{1a}
\frac{1}{M_*^2}\int{\sqrt{-g} J^\mu \partial_\mu (R) d^4x}
\end{equation}

In Eq. \eqref{1a}, the parameter $M_*$ represents the cutoff scale of the effective theory, whereas $R$, $J^\mu$, and $g$ stand for the  Ricci scalar, baryonic matter current, and the trace of the metric tensor respectively. The baryon to entropy ratio $\frac{n_B}{s} \propto \dot{R}$, in the case of flat FRW Universe, where the dot represents cosmic time derivative. In the situation of a radiation-dominated era with $\omega=\frac{1}{3}$, the net baryon asymmetry generated by Eq.\eqref{1a} is zero.
 In the past few years, a number of authors have studied the mysterious concept of Baryogenesis within the context of modified gravity. In \cite{Lamb1,Ramo}, $f(R)$ gravity theories are addressed in terms of gravitational baryogenesis, whereas in \cite{Odin} Gauss-Bonnet gravity, in \cite{Bent} Gauss-Bonnet braneworld cosmology, in \cite{Oiko} $f (T)$ gravity, in \cite{Bhat} $f(P)$ gravity, teleparallel gravity \cite{Bhat2}, in \cite{Noza,Baff,Saho} $f (R,T)$ gravity, in \cite{Sale} $f(R,T,X)$ gravity, in \cite{Azha} $f(T,B)$ gravity, and in \cite{Bhat1} $f(Q,T)$ gravity, etc. 
 The objective of this work is to look into the framework of gravitational baryogenesis in the $f(R,L_m)$ gravity theory, which is an extension of the $f(R)$ gravity theory \cite{Star}, that includes an explicit coupling of Rcicci scalar $R$ and matter Lagrangian $L_m$.  This Theory was proposed by Harko and Lobo \cite{Hark} as well as several works with interesting results have been found \cite{Hark1,Jayb,Labo,Labo1,Jayb1}. The following is how the present manuscript is organized: Section \ref{sec2} includes an overview of $f(R,L_m)$ gravity. In Section \ref{sec3}, we will provide some essential baryogenesis factors before investigating gravitational baryogenesis in $f(R,L_m)$ gravity and then we explain baryogenesis in $f(R,L_m)$ gravity in detail and imply generating observationally acceptable baryon to entropy ratios for $f(R,L_m)$ gravity model and we will also go over the generalized form of baryogenesis that applies to assuming gravity model. In the final section \ref{sec4}, we will discuss the conclusions of the present work.

\section{Overview of $f(R,L_m)$ Gravity}\label{sec2}

\justify

Consideration of the overall action in modified $f(R,L_m)$ gravity \cite{Hark}

\begin{equation}\label{2a}
S= \int{f(R,L_m)\sqrt{-g}d^4x} 
\end{equation}

Where $R$, $L_m$ stand for the Ricci scalar curvature and the matter Lagrangian. 
 By varying the action \eqref{2a} with respect to the metric tensor $g_{\mu\nu}$, we were able to derive the field equation governing the dynamics of gravitational interactions.
 
\begin{equation}\label{2b}
f_R R_{\mu\nu} + (g_{\mu\nu} \square - \nabla_\mu \nabla_\nu)f_R - \frac{1}{2} (f-f_{L_m}L_m)g_{\mu\nu} = \frac{1}{2} f_{L_m} T_{\mu\nu}
\end{equation}

Here $f_R \equiv \frac{\partial f}{\partial R}$, $f_{L_m} \equiv \frac{\partial f}{\partial L_m}$, and $T_{\mu\nu}$ represents the stress-energy tensor for the ideal fluid, which is described by 

\begin{equation}\label{2c}
T_{\mu\nu} = \frac{-2}{\sqrt{-g}} \frac{\delta(\sqrt{-g}L_m)}{\delta g^{\mu\nu}}
\end{equation}

It further takes the covariant derivative in equation \eqref{2b} to yield the following result:

\begin{equation}\label{2d}
\nabla^\mu T_{\mu\nu} = 2\nabla^\mu ln(f_{L_m}) \frac{\partial L_m}{\partial g^{\mu\nu}}
\end{equation}

Now, we think of a spatially flat FLRW metric as

\begin{equation}\label{2e}
ds^2= -dt^2 + a^2(t)[dx^2+dy^2+dz^2]
\end{equation}

\justify where, $ a(t) $ is the cosmic scale factor. So the Ricci scalar  produced pertaining to the line element \eqref{2e} is

\begin{equation}\label{2f}
R= 6 \frac{\ddot{a}}{a}+ 6 \bigl( \frac{\dot{a}}{a} \bigr)^2 = 6 ( \dot{H}+2H^2 )
\end{equation}

\justify Overall, we get the modified Friedman equations that characterize the dynamics of the universe in $f(R,L_m)$ gravity by employing \eqref{2e} in \eqref{2b} \cite{Jayb},

\begin{equation}\label{2g}
3H^2 f_R + \frac{1}{2} \left( f-f_R R-f_{L_m}L_m \right) + 3H \dot{f_R}= \frac{1}{2}f_{L_m} \rho 
\end{equation}
and
\begin{equation}\label{2h}
\dot{H}f_R + 3H^2 f_R - \ddot{f_R} -3H\dot{f_R} + \frac{1}{2} \left( f_{L_m}L_m - f \right) = \frac{1}{2} f_{L_m} p
\end{equation} 
where  $H=\frac{\dot{a}}{a}$, the Hubble parameter, and the dot stand for the derivative with respect to cosmic time $t$. The matter density and pressure are denoted by $\rho$ and $p$ in the equations above.

\section{ Baryogenesis in $f(R,L_m)$}\label{sec3}
\justifying
We shall demonstrate how $f(R,L_m)$ gravity addresses the gravitational baryogenesis difficulty in the cosmos in this part. A crucial parameter to figure out asymmetry is known as BAF given by

\begin{equation}\label{3a}
    \eta _B =\frac{n_B - \Bar{n}_B}{s}
\end{equation}

where $n_B$ is the baryon number and $\bar{n}_B$ is the antibaryon number, and s is the entropy of the universe. Observational evidence like the BBN \cite{Burl} and CMB \cite{Benn} confirm that the restriction on this BAF is $\frac{n_B}{s}\simeq 9\times10^{-11}$. For the $f(R,L_m)$ gravity, we take into consideration an interaction term that violates CP and is produced by the baryon asymmetry of the universe of the form

\begin{equation}\label{3b}
\frac{1}{M_*^2}\int{\sqrt{-g} J^\mu \partial_\mu (R+L_m) d^4x}
\end{equation}

As a result, for an interaction of \eqref{3b} that violates CP, the resulting baryon to entropy ratio in $f(R,L_m)$ gravity is as follows:

\begin{equation}\label{3c}
\frac{n_B}{s}\simeq -\frac{15 g_B }{4\pi ^2  g_{*s} }\frac{(\dot{R}+\dot{L_m})}{M^2 _* T_D}
\end{equation}

In \eqref{3c}, $g _B$ is the total number of intrinsic degrees of freedom of baryons, $g_*s$ is the total number of degrees of freedom of massless particles, and the critical temperature is $T_D$ is the temperature of the universe when all interactions that cause baryon asymmetry to begin. We will assume that a thermal equilibrium exists, with energy density being proportional to temperature $T$ as

\begin{equation}\label{3d}
\rho(T)=\frac{\pi^2}{30} g_{*s} T^4
\end{equation}

Inside the framework of Einstein's general theory of relativity, suppose the universe's matter content is a perfect fluid with the constant equation of state parameter $\omega=\frac{p}{\rho}$ and the Ricci scalar $R$, as

\begin{equation}\label{3e}
 R=-8\pi G (1-3\omega)\rho   
\end{equation}

In General Relativity (GR), if the universe is filled with radiation, the baryon number to entropy ratio equals zero. For the other content of the matter, this result differs from zero. However, in $f(R,L_m)$ gravity theories, a net baryon asymmetry may be generated during the radiation-dominated era. To accomplish this, we focus on a specific $f(R,L_m)$ model to describe how we can recover the baryogenesis epoch with this model. We calculate the baryon-to-entropy ratio for the model by imagining a universe full of dark energy and perfect fluid with the constant equation of state parameter $\omega=\frac{P}{\rho}$.

\subsection*{3.1\, \textbf{The perfect fluid with $f(R,L_m)$ gravity}}

The recently proposed model we are using in this section was presented in Ref. \cite{Jayb}. In our work, we borrowed this model and its correlating field equations from this paper and utilized them to explain baryogenesis. That is

\begin{equation}\label{3f}
  f(R,L_m) = \frac{R}{2} + L_m ^{\alpha} + \zeta  
\end{equation}

where $\alpha$ and $\zeta$  are model parameters. Then, for this specific $f(R,L_m)$ model with $L_m=\rho$ \cite{Hark1}, the Friedmann equations \eqref{2g} and \eqref{2h} are transformed into

\begin{equation}\label{3g}
3H^2 = (2\alpha-1) \rho ^{\alpha}-\zeta
\end{equation}
and
\begin{equation}\label{3h}
-2\dot{H}-3H^2 =(\alpha p+ (1-\alpha)\rho)\rho^{\alpha-1}+\zeta
\end{equation} 

One can obtain the following energy balance equation by taking the trace of the field equations

\begin{equation}\label{3i}
\alpha \dot{\rho}+ 3H \rho = 0
\end{equation}

To continue, we will presume that the scale factor adapts as a power-law $a(t)= B t^{\beta}$, where $\beta = \frac{2}{3(1+\omega)}$, and $B$ is a constant parameter \cite{Odin}. Then the Hubble parameter $H(t)$ and the energy density $\rho(t)$ expressions for this model are as follows

\begin{equation}\label{3j}
H(t) =  \frac{\beta }{t}
\end{equation}

\begin{equation}\label{3k}
\rho(t) =  \big[\frac{3\beta ^2 +\zeta t^2}{(2\alpha-1)t^2}  \big]^{\frac{1}{\alpha}}
\end{equation}

Equating Eqs. \eqref{3k} and \eqref{3d}, we obtain the decoupling time $t_D$ as a function of the decoupling temperature $T_D$ written as

\begin{equation}\label{3l}
t_D=\big[\frac{3\beta ^2}{(2\alpha-1)(\frac{\pi ^2}{30} g_{*s} T_D ^4)^\alpha -\zeta}\big]^\frac{1}{2}
\end{equation}

Using \eqref{3l}, we retrieve a final expression for the baryon-to-entropy ratio for the current $f(R, L_m)$ model

\begin{widetext}
\begin{multline}\label{3m}
\frac{n_B}{s}\simeq \left(\frac{-15 g_B}{4\pi ^2 g_{*s}M^2 _*  T_D} \right)\left(\frac{\left[(2\alpha-1)(\frac{\pi ^2}{30} g_{*s}T_D ^4 )^\alpha -\zeta \right]^\frac{3}{2}}{\sqrt{27} \beta ^3}\right) \left(12(\beta-2\beta ^2)-\frac{ 6\beta \left(\frac{\pi ^2}{30} g_{*s}T_D ^4 \right)^{1-\alpha}}{\alpha (2\alpha-1)}\right)
\end{multline}
\end{widetext}

In the radiation-dominated phase, $\beta=\frac{1}{2}$.
Hence, \eqref{3m} reduces to

\begin{equation}\label{3n}
\frac{n_B}{s}\simeq \left(\frac{-15 g_B}{4\pi ^2 g_{*s}M^2 _*  T_D} \right)\left(\frac{8\left[(2\alpha-1)(\frac{\pi ^2}{30} g_{*s}T_D ^4 )^\alpha -\zeta \right]^\frac{3}{2}}{\sqrt{27} }\right) \left(-\frac{ 3 \left(\frac{\pi ^2}{30} g_{*s}T_D ^4 \right)^{1-\alpha}}{\alpha (2\alpha-1)}\right)
\end{equation}

 As shown in Eq.\eqref{3n}, the resulting baryon to entropy ratio is non-zero. The ratio \eqref{3n} can be adjusted to satisfy the observational constraints depending on the matter content, but the most interesting feature of a perfect fluid dominated Universe for the case of $f(R,L_m)$ gravity baryogenesis is that the ratio is non-zero in the radiation dominated case.

\begin{figure}[H]
\centering
\includegraphics[scale=1.0]{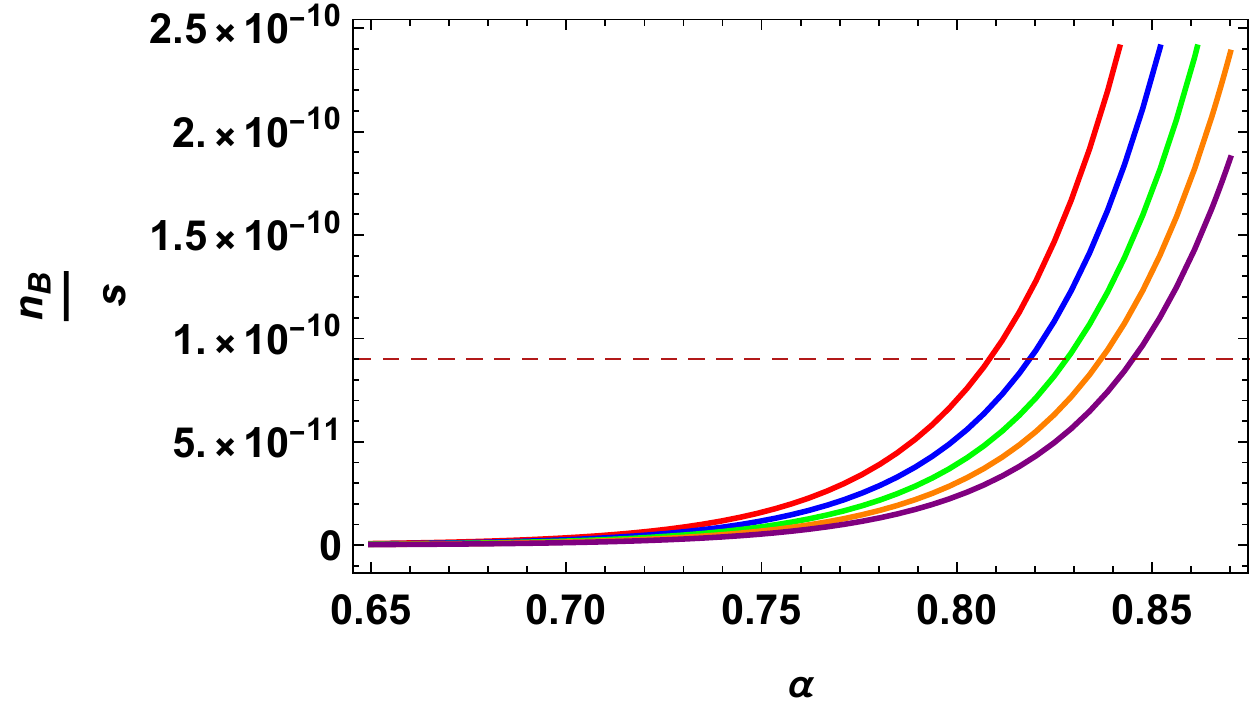}
\caption{The baryon to entropy ratio for the model $f (R,L_m)=\frac{R}{2} + L_m ^\alpha + \zeta$. The graphs are plotted for $\alpha$, for the varying values of $\beta$, $\beta=0.55$ (Red), $\beta=0.60$ (Blue), $\beta=0.65$ (Green), $\beta=0.70$ (Orange), $\beta=0.75$ (Purple), and $\zeta=2$. The dashed line represents the observational value.}\label{1}
\end{figure}

Substituting $g_{*s} = 106$, $g_B =1$, $T_D = 2\times 10^{12} GeV$ and $M_* = 2 \times 10^{16} GeV$ \cite{Lamb}, with model parameters $\alpha=0.79$ and $\zeta=2$  in Eq. \eqref{3n} the resultant baryon to entropy ratio reads $\frac{n_B}{s}\simeq 7.28749 \times 10^{-11}$ which is in excellent agreement with observations. The intersections of the curves in \textbf{Fig} \ref{1} with the curve that reveals the observational value (dashed line) are in fine contract value of baryon to entropy ratio for particular values of $\alpha$ about including $0.75$ and $0.85$. Interestingly, when the parameter $0.5\leq \alpha \leq 0.69$, thus every curve tends to zero, which is consistent with theoretical results.

\subsection*{3.2 \, \textbf{Generalized Gravitational Baryogenesis}}
\justify
We will now attempt to investigate the effects of a more comprehensive and generalized CP-violating interaction proportional to $\partial _\mu (f(R,L_m))$ rather than $\partial _\mu (R+L_m)$ in attempting to address the baryon asymmetry of the Universe for the chosen $f(R,L_m)=\frac{R}{2} + L_m ^\alpha + \zeta$  model. In $f(R,L_m)$ gravity, we can express the generalized CP-violating interaction as \cite{Noza}
%We will now define a more generalized and complete baryogenesis interaction proportional to  $\partial _\mu f(R,L_m)$. The interaction that violates the CP then reads \cite{Noza}

\begin{equation}\label{4a}
\frac{1}{M_*^2}\int{\sqrt{-g} J^\mu \partial_\mu (f(R,L_m)) d^4x}
\end{equation}

The resulting baryon-to-entropy ratio for \eqref{4a} is

\begin{equation}\label{4b}
\frac{n_B}{s}\simeq \frac{-15 g_B (\dot{R} f_R+\dot{L_m}f_{L_m})}{4 g_{*s}M^2 _* \pi ^2 T_D}
\end{equation}

%\textbf{For model $f(R,L_m)=\frac{R}{2} + L_m ^\alpha + \zeta$ we obtain $\dot{R} f_R+\dot{L_m}f_{L_m}=\frac{1}{2}\dot{R}+\alpha\dot{L_m}L_m ^{\alpha-1} $. 
Substituting \eqref{3e}, \eqref{3f}, \eqref{3j}, and \eqref{3l} in \eqref{4b}, we get the baryon to entropy ratio as

%We get the baryon to entropy ratio by plugging all the values into equation \eqref{4b}

\begin{widetext}
\begin{multline}\label{4c}
\frac{n_B}{s}\simeq \left(\frac{-15 g_B}{4\pi ^2 g_{*s}M^2 _*  T_D} \right)\left(\frac{\left[(2\alpha-1)(\frac{\pi ^2}{30} g_{*s}T_D ^4 )^\alpha -\zeta \right]^\frac{3}{2}}{\sqrt{27} \beta ^3}\right) \left(6(\beta-2\beta ^2)-\frac{ 6\beta}{(2\alpha-1)}\right)
\end{multline}
\end{widetext}

We finally obtain, as discussed in the previous section for a radiation-dominated universe by setting $\beta=\frac{1}{2}$, in Eq. \eqref{4c}

\begin{equation}\label{4d}
\frac{n_B}{s}\simeq \left(\frac{-15 g_B}{4\pi ^2 g_{*s}M^2 _*  T_D} \right)\left(\frac{8\left[(2\alpha-1)(\frac{\pi ^2}{30} g_{*s}T_D ^4 )^\alpha -\zeta \right]^\frac{3}{2}}{\sqrt{27}}\right) \left(-\frac{ 3}{4(2\alpha-1)}\right)
\end{equation}

By substituting  $M_*$, $g_{*s}$, $g_B$, $T_D$ as before, $\alpha=0.93$, and $\zeta=2$, the obtained baryon to entropy ratio is $\frac{n_B}{s}\simeq 7.01\times10^{-11} $,  which is also in excellent agreement with observations.

\begin{figure}[H]
\centering
\includegraphics[scale=1.0]{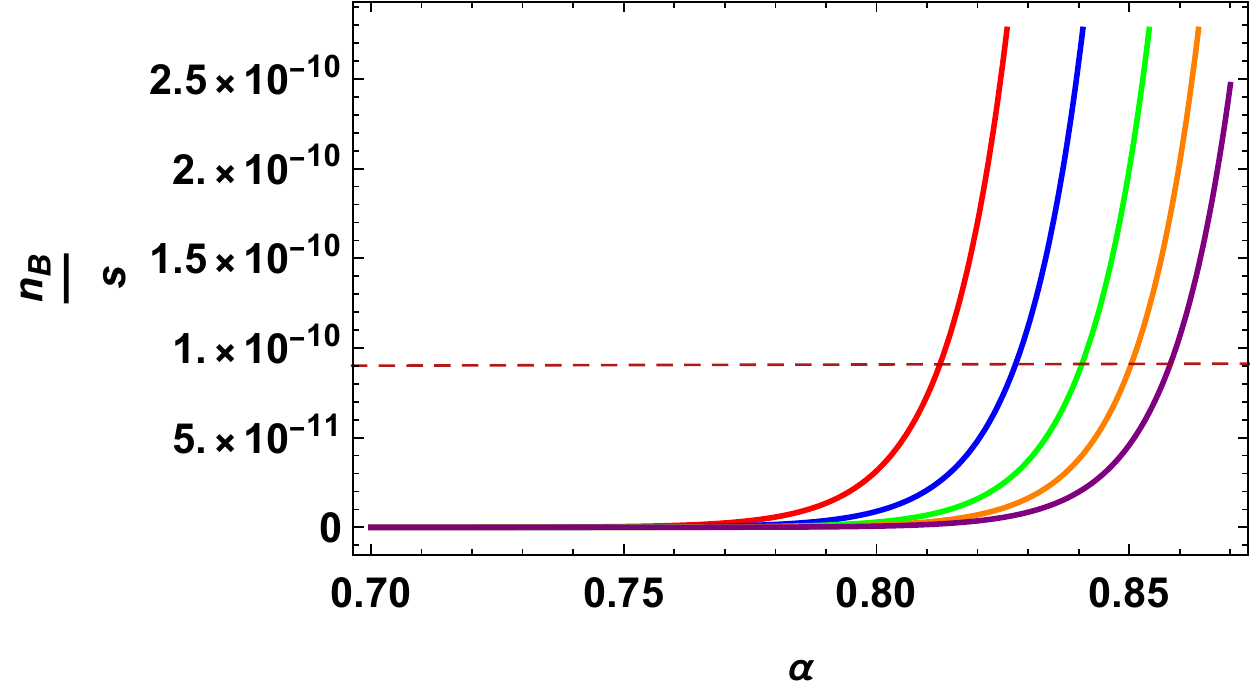}
\caption{The baryon to entropy ratio for the model $f (R,L_m)=\frac{R}{2} + L_m ^\alpha + \zeta$. The graphs are plotted for $\alpha$, for the varying values of $\beta$, $\beta=0.018$ (Red), $\beta=0.026$ (Blue), $\beta=0.036$ (Green), $\beta=0.046$ (Orange), $\beta=0.056$ (Purple), and $\zeta=2$. The dashed line represents the observational value.}\label{2}
\end{figure} 
 
%By substituting  $M_*$, $g_{*s}$, $g_B$, $T_D$ as before, $\alpha=0.93$, and $\zeta=2$, the obtained baryon to entropy ratio is $\frac{n_B}{s}\simeq 7.01\times10^{-11} $,  which is also in excellent agreement with observations.
We show  $\frac{n_B}{s}$ for the generalised baryogenesis interaction as a function of $\alpha$ in \textbf{Fig} \ref{2}.
The intersections of the curves in \textbf{Fig} \ref{2} with the curve that reveals the observational value (dashed line) are in fine contract value of baryon to entropy ratio for particular values of $\alpha$ about including $0.79$ and $0.86$. Interestingly, when the parameter $0.6\leq \alpha \leq 0.79$, thus every curve tends to zero, which is consistent with theoretical results. Acceptable baryon to entropy values was obtained by $f (R,L_m)=\frac{R}{2} + L_m ^\alpha + \zeta$, which resulted in physically acceptable baryon to entropy ratios. Thus, in $f(R,L_m)$ gravity, the problem of baryogenesis can be rectified.

\section{ Conclusions}\label{sec4}
\justify

The paper investigates the methodology of gravitational baryogenesis from the perspective of $f(R,L_m)$ gravity theory. We evaluate the baryon to entropy ratio for specific  $f(R,L_m)=\frac{R}{2}+L_m ^\alpha+\zeta$ model based on the CP-violating interaction that will produce the Universe's baryon asymmetry and assessing the Universe's matter content as a perfect fluid with a constant equation of state parameter $\omega$. In comparison to GR, we demonstrate that its baryon-to-entropy ratio in a radiation-dominated era is nonzero for our model. Then we find the baryon-to-entropy ratio for a radiation-dominated universe $\beta=\frac{1}{2}$, and when the Universe is filled with perfect fluid, and cosmic dynamics are governed by the $f(R,L m)$ theory of gravity. We determined the baryon-to-entropy ratio with specific model $\frac{n_B}{s}\simeq 7.28749 \times 10^{-11}$, which is in excellent agreement with the observational value of $9\times 10^{-11}$.
We assumed a scale factor of the form $a(t)= B t^{\beta}$, of $\beta = \frac{2}{3(1+\omega)}$, and $B$ is a constant parameter for this work. Because $\beta=\frac{1}{2}$ were the appropriate values required to obtain a viable baryon-to-entropy ratio, we assert that for such a scale factor, the power law part predominated at early times, which is consistent with observations \cite{Ries,Perl}.
Finally, we conclude our research by investigating a more complete and generalized baryogenesis interaction that is proportional to $\partial _\mu f(R,L_m)$. In this kind of interaction, our model produced a theoretical value in leading order of $\frac{n_B}{s}\simeq 7.01\times10^{-11} $, which is close to the observed value as well as the value obtained in the first case.

\section*{Data Availability Statement}
There are no new data associated with this article.

\section*{Acknowledgements}

L.V.J. acknowledges the University Grant Commission (UGC), Govt. of India, New Delhi, for awarding JRF (NTA Ref. No.: 191620024300). PKS acknowledges Science and Engineering Research Board, Department of Science and Technology, Government of India for financial support to carry out Research project No.: CRG/2022/001847 and IUCAA, Pune, India for providing support through the visiting Associateship program. We are very much grateful to the honorable referee and to the editor for the illuminating suggestions that have significantly improved our work in terms of research quality, and presentation.

\end{document}